\documentclass[12pt]{article}
\usepackage{amsmath,amssymb,bm,epsfig,color}
\renewcommand{\thefootnote}{\fnsymbol{footnote}}

\begin{document}

\title{
\begin{flushright}
\begin{minipage}{0.2\linewidth}
\normalsize
EPHOU-17-015 \\
WU-HEP-17-015 \\*[50pt]
\end{minipage}
\end{flushright}
{\Large \bf 
Radiative K\"ahler moduli stabilization
\\*[20pt]}}

\author{Tatsuo~Kobayashi,$^{1,}$\footnote{
E-mail address: kobayashi@particle.sci.hokudai.ac.jp}\ \
Naoya Omoto,$^{1,}$\footnote{
E-mail address: omoto@particle.sci.hokudai.ac.jp}\ \
Hajime~Otsuka,$^{2,}$\footnote{
E-mail address: h.otsuka@aoni.waseda.jp}\\
\ and \
Takuya~H.~Tatsuishi$^{1}$\footnote{
E-mail address: t-h-tatsuishi@particle.sci.hokudai.ac.jp}
\\*[20pt]
$^1${\it \normalsize 
Department of Physics, Hokkaido University, Sapporo 060-0810, Japan}  \\
$^2${\it \normalsize 
Department of Physics, Waseda University, 
Tokyo 169-8555, Japan
} \\*[50pt]}

\date{
\centerline{\small \bf Abstract}
\begin{minipage}{0.9\linewidth}
\medskip 
\medskip 
\small 
We propose a new type of K\"ahler moduli stabilization mechanisms 
in type IIB superstring theory on Calabi-Yau manifolds 
with the positive Euler number. 
The overall K\"ahler modulus can be perturbatively stabilized by 
radiative corrections due to sparticles.
Its minimum is the anti-de Sitter vacuum, where supersymmetry is 
broken.
We can uplift it to the de Sitter vacuum by introducing anti-D-branes, 
keeping the modulus stabilized.
Although our numerical results depend on the choice of the cutoff scale and 
degeneracies of sparticles, at any rate there exist the parameter spaces where the masses of 
Kaluza-Klein and stringy modes are larger than the cutoff scale. 
Furthermore, this stabilization scenario predicts an ultralight axion. 
\end{minipage}
}

\begin{titlepage}
\maketitle
\thispagestyle{empty}
\end{titlepage}

\renewcommand{\thefootnote}{\arabic{footnote}}
\setcounter{footnote}{0}

\tableofcontents

\section{Introduction}
The moduli fields appear ubiquitously in four-dimensional 
low-energy effective field theory derived from superstring theory 
on six-dimensional compact space.
These fields correspond to a geometrical character of compact space.
Unless these moduli fields are stabilized at a high scale, 
it will lead to the fifth force. 
Hence, moduli stabilization is one of the major topics in the string 
cosmology and phenomenology. 

So far, moduli stabilization has been well studied in the type IIB superstring theory 
on Calabi-Yau (CY) manifolds.
The closed string moduli are categorized into the dilaton $S$, complex 
structure moduli $U$ and  K\"ahler moduli $T$.\footnote{In this paper, we do not 
discuss the stabilization of open string moduli. For discussion of 
open string moduli in F-theory context, see e.g. Ref.~\cite{Dasgupta:1999ss,Honma:2017uzn}, where the open string moduli are identified 
with the complex structure moduli of CY fourfold.} 
Here, we denote the numbers of complex structure moduli and K\"ahler moduli by 
$h^{2,1}$ and $h^{1,1}$, which correspond to the numbers of three-cycles and four-cycles of 
CY, respectively. 
Three-form fluxes in the type IIB superstring theory can stabilize the dilaton $S$ 
and all the complex structure moduli $U$ at the compactification scale~\cite{Giddings:2001yu} and 
they generically generate the nonvanishing flux-induced superpotential $W_0=\langle W_{\rm flux}\rangle$~\cite{Gukov:1999ya}. 
The remaining K\"ahler moduli are stabilized at an anti-de Sitter minimum in the Kachru-Kallosh-Linde-Trivedi (KKLT) 
scenario~\cite{Kachru:2003aw} and 
LARGE volume scenario (LVS)~\cite{Balasubramanian:2005zx} using the non-perturbative superpotential for the K\"ahler 
moduli. Such an anti-de Sitter minimum is uplifted to a metastable de Sitter minimum by introducing  anti-D3 branes~\cite{Kachru:2003aw}. 
The $F$-term uplifting scenario is another way to realize the de Sitter minimum \cite{Lebedev:2006qq,Dudas:2006gr}.
In the KKLT scenario, the stabilization of K\"ahler moduli is achieved by tuning $|W_0| \ll M_{\rm Pl}^3$, where $M_{\rm Pl}$ is 
the reduced Planck mass.
On the other hand, even if $|W_0| \sim {\cal O}(M_{\rm Pl}^3)$, the LVS works for CY manifolds with the negative Euler number, 
$\chi=2(h^{1,1}-h^{2,1})<0$, i.e., $h^{2,1} > h^{1,1}> 1$.\footnote{For the LVS with zero or positive Euler number, see 
Ref.~\cite{Cicoli:2012fh}.} 
Note that stringy modes and Kaluza-Klein modes should be 
sufficiently heavier than all the moduli fields to justify description of the low-energy effective field theory.

In this paper, along the line of Ref.~\cite{Kobayashi:2017aeu}, 
we propose a new type of K\"ahler moduli stabilization mechanisms in type IIB superstring 
theory on CY orientifolds with the positive Euler number $\chi>0$, namely $h^{1,1}>h^{2,1}>1$. 
In Ref.~\cite{Berg:2005yu}, it was discussed that the overall vomlume modulus can be 
stabilized perturbatively by one-loop corrections to the K\"ahler potential.
In that scenario, one needs a certain amount of fine-tuning of complex structure moduli to 
realize the K\"ahler moduli stabilization.
 In contrast to Ref.~\cite{Berg:2005yu}, 
here we focus on the radiative corrections due to the sparticles living on D7-branes wrapping a divisor of CY. 
Such corrections generate the potential of overall volume modulus through the Coleman-Weinberg potential~\cite{Coleman:1973jx}. 
Since the soft masses are functions of only the overall volume modulus, the overall 
K\"ahler modulus can be perturbatively stabilized at a sufficiently large 
volume region without tuning $W_0$. In this large volume region, 
the string axion associated with the overall volume modulus 
remains light and it could be a candidate of dark matter. 
Other K\"ahler moduli could be stabilized by the non-perturbative effects and/or 
moduli-dependent $D$-terms. (See, e.g. Refs.~\cite{Jockers:2005zy}.) 
We find that stringy and Kaluza-Klein modes are 
sufficiently heavier than sparticles and K\"ahler moduli at the anti-de Sitter minimum. 
We can uplift the anti-de Sitter vacuum to the de Sitter vacuum by 
introducing anti-D branes, and such uplifting does not change the behavior 
of the moduli stabilization.

The remaining paper is organized as follows. 
In Sec.~\ref{sec:2}, we discuss the stabilization of overall K\"ahler modulus without uplifting terms 
in type IIB superstring theory on CY orientifolds with D$7$-branes. 
We add the uplifting term in the setup of Sec.~\ref{sec:3} to achieve a tiny cosmological 
constant.
We study the modulus potential with the uplifting terms analytically and numerically 
in Secs.~\ref{sec:3_1} and~\ref{sec:3_2}, respectively. Other K\"ahler moduli 
can be stabilized by two scenarios.
One is due to moduli-dependent $D$-terms as discussed in Sec.~\ref{sec:3_3}, and 
the other is due to non-perturbative effects as discussed in Sec.~\ref{sec:3_4}.
Also, in Sec.~\ref{sec:3_5}, we show that 
the non-perturbative effects generate the mass of ultralight axion associated with 
the volume modulus. 
Finally, Sec.~\ref{sec:con} is devoted to the conclusion.

\section{K\"ahler moduli stabilization without uplifting}
\label{sec:2}
\subsection{Setup}
Let us consider the stabilization of closed string moduli on the basis of type IIB superstring theory on CY orientifolds 
with D$7$-branes. 
For a general class of CY threefolds, the total K\"ahler potential at the 
leading order of $\alpha^\prime$ is obtained through Kaluza-Klein 
reduction of type IIB supergravity action~\cite{Becker:2002nn,Grimm:2004uq},
\begin{align}
K&=-2M_{\rm Pl}^2\ln \left({\cal V}+\frac{\xi}{2}\right)+M_{\rm Pl}^2K(S,U) +Z_{a\bar{a}}^{(i)}|Q_a^{(i)}|^2,
\label{eq:K}
\end{align}
where ${\cal V}$ is the volume of CY manifold in Einstein frame measured by the string length $l_s=2\pi \sqrt{\alpha^\prime}$, 
$K(S,U)$ represents the K\"ahler potential of dilaton $S$ and complex structure moduli $U$. The 
leading $\alpha^\prime$-correction is characterized by $\xi=-\frac{\zeta(3)\chi}{2(2\pi)^3g_s^{3/2}}$ where $\chi$ 
is the Euler number of CY and $g_s$ is the string coupling~\cite{Becker:2002nn}.
Here, $Q_a^{(i)}$ are the matter fields living on D$7$-branes wrapping the divisor $D_i\in H_4({\rm CY},\mathbb{Z})$ 
and their K\"ahler metrics $Z_{a\bar{a}}^{(i)}$ are 
given by 
\begin{align}
Z_{a\bar{a}}^{(i)}=(T_i+\bar{T}_i)^{-n^a_i} ,
\end{align}
with $n_i^a$ being the modular weights and $T_i=\tau_i +i \sigma_i$ with $i\in \{1,2,\cdots, h^{1,1}_+\}$ 
denotes the K\"ahler modulus.\footnote{In this paper, we consider the orientifold projection to realize $h^{1,1}_-=0$, 
namely $h^{1,1}=h^{1,1}_+$.} 
Also, the gaugino fields couple to the K\"ahler moduli through the gauge kinetic function
\begin{align}
f=\frac{T_i}{2\pi}.
\end{align}
To stabilize the complex structure moduli $U$ and dilaton $S$, we consider the 
following superpotential induced by imaginary self-dual three-form fluxes~\cite{Gukov:1999ya},
\begin{align}
W=W_{\rm flux}(S,U).
\end{align}

By the flux-induced superpotential, all the complex structure moduli and 
dilaton fields are stabilized at the compactification scale~\cite{Giddings:2001yu}. 
When $W_0=\langle W_{\rm flux}\rangle\neq 0$ at the minima of the complex 
structure moduli and dilaton, the scalar potential is nonvanishing
because of the breaking of the so-called no-scale structure~\cite{Becker:2002nn}
\begin{align}
V_{\alpha^\prime}=e^{\langle K(S,U)\rangle}\left(K^{T_i\bar{T}_j}K_{T_i}K_{\bar{T}_j}-3\right)\frac{|W_0|^2}{M_{\rm Pl}^2}\simeq 
e^{\langle K(S,U)\rangle}\frac{3\xi}{4{\cal V}^3}\frac{|W_0|^2}{M_{\rm Pl}^2},
\end{align}
where $K_{T_i}=\partial_{T_i} K$ and 
$K^{T_i\bar{T}_j}$ is the inverse of K\"ahler metric $K_{T_i\bar{T}_j}=\partial_{T_i}\partial_{\bar{T}_j}K$. 
The nonvanishing $F$-term of the K\"ahler modulus 
\begin{align}
F^{T_i}=-e^{K/(2M_{\rm Pl}^2)}K^{T_i\bar{T}_j} K_{\bar{T}_j}\frac{\bar{W}_0}{M_{\rm Pl}},
\end{align}
generates the soft scalar masses and 
gaugino masses~\cite{Kaplunovsky:1993rd},
\begin{align}
m_a^2&=\frac{V_{\alpha^\prime}}{M_{\rm Pl}^2}+m_{3/2}^2 -(\partial_{T_i}\partial_{\bar{T}_j}\ln(Z_{a\bar{a}}^{(i)}))
\frac{F^{T_i}\bar{F}^{T_j}}{M_{\rm Pl}^2}
=\frac{V_{\alpha^\prime}}{M_{\rm Pl}^2}+m_{3/2}^2-\frac{n^a_i}{M_{\rm Pl}^2}\left|\frac{F^{T_i}}{T_i+\bar{T}_i}\right|^2 
\nonumber\\
&\simeq \frac{V_{\alpha^\prime}}{M_{\rm Pl}^2}+(1-n_i^a)m_{3/2}^2 
\simeq (1-n_i^a)m_{3/2}^2,
\nonumber\\
M_f&=\frac{F^{T_i}}{M_{\rm Pl}}\partial_{T_i} \ln {\rm Re}(f)=M_{\rm Pl}^{-1}\frac{F^{T_i}}{T_i+\bar{T}_i}
\simeq 
\frac{e^{\langle K(S,U)\rangle/2}\bar{W}_0 }{{\cal V}M_{\rm Pl}^2} \simeq m_{3/2},
\label{eq:softmass}
\end{align}
with
\begin{align} 
m_{3/2}=e^{K/(2M_{\rm Pl}^2)}W_0M_{\rm Pl}^{-2}\simeq \frac{e^{\langle K(S,U)\rangle/2}W_0 }{{\cal V}M_{\rm Pl}^2},
\end{align}
evaluated in the large volume region.\footnote{We take $W_0$ as a real constant for simplicity and 
discuss the case with $n^a_i=0$ later.} 
Here, we evaluate $\frac{F^{T_i}}{T_i+\bar{T}_i}\simeq -e^{K/(2M_{\rm Pl}^2)}\frac{\bar{W}_0}{M_{\rm Pl}}
\simeq -m_{3/2}M_{\rm Pl}$, which is   
satisfied in a general class of CY threefolds with a sufficiently large volume. 
When these sparticles contribute to the radiative corrections, 
the 1-loop Coleman-Weinberg (CW) potential is given by
\begin{align}
V_{\rm CW}&=\frac{1}{32\pi^2}\int^{\Lambda^2} dk^2 k^2 {\rm STr}\ln (k^2+M^2)
\nonumber\\
&=\frac{1}{32\pi^2}\left\{
\Lambda^2 {\rm STr}(M^2) + \frac{1}{2}{\rm STr}M^4 \biggl[ \ln \left(\frac{M^2}{\Lambda^2}\right)
-\frac{1}{2}+{\cal O}\left(\frac{M^2}{\Lambda^2}\right)\biggl]\right\}
\nonumber\\
&\simeq \frac{c_1}{32\pi^2}
\Lambda^2 m_{3/2}^2
+\sum_a\frac{c_b^a}{64\pi^2}m_{a}^4\biggl[ \ln\left(\frac{m_{a}^2}{\Lambda^2}\right)-\frac{1}{2}\biggl]
+\frac{-2c_f-4}{64\pi^2}m_{3/2}^4 \biggl[\ln\left(\frac{m_{3/2}^2}{\Lambda^2}\right)-\frac{1}{2}\biggl]
\nonumber\\
&+\frac{32}{64\pi^2}m_{3/2}^4 \biggl[\ln\left(\frac{4m_{3/2}^2}{\Lambda^2}\right)-\frac{1}{2}\biggl],
\end{align}
where $M$ characterizes the mass matrix for canonically normalized bosons and fermions, and 
\begin{align}
c_1=\sum_a c_b^a(1-n_i^a)-2c_f+4,
\label{eq:c1}
\end{align}
with $c_b^a$ and $c_f$ being the multiplicities of bosons $Q_a^{(i)}$\footnote{We use the same notation of 
chiral superfields $Q_a$ and their lowest components.} and fermions, respectively. 
Here, we take the limit $M^2/\Lambda^2 \ll 1$, where  our description of low-energy effective theory 
is valid and 
the supertrace is defined as
\begin{align}
{\rm Str}f(M^2)={\rm tr}f(m_a^2)-2{\rm tr}f(M_f^2)-4f(m_{3/2}^2)+2f(4m_{3/2}^2),
\label{eq:supertrace}
\end{align}
for an arbitrary function $f$. We consider the contribution of 
ghosts in the gauge $\sum_{\mu=0}^3\gamma^\mu\psi_\mu=0$, 
where $\psi_\mu$ and $\gamma^\mu$ are the gravitino and four-dimensional gamma-matrices, respectively. (For more details, see Refs.~\cite{Barbieri:1883bv}.) 
Obviously, there is no contribution from sfermion fields with $n^a_i=1$.

For example, in the minimal supersymmetric standard model (MSSM) with three generations of 
right-handed (s)neutrinos, 
those multiplicities become
\begin{align}
c_b^{({\rm MSSM})}=\sum_a c_b^a=49+3=52,\qquad
c_f^{({\rm MSSM})}=12.
\end{align}
On the other hand, when the visible sector consists of the MSSM with singlets and multi-Higgs doublets, 
the multiplicities are given by 
\begin{align}
c_b=c_b^{({\rm MSSM})}+4(n_H-1)+n_S,\qquad
c_f^{({\rm MSSM})}=12,
\label{eq:cb}
\end{align}
where $n_H$ and $n_S$ denote the numbers of multi-Higgs doublets and singlet fields. 
Note that the following moduli stabilization is also applicable to the scenario 
where sparticles living on hidden D$7$-branes contribute to the radiative corrections. 
If there exist multiple D$7_i$-branes wrapping the divisor $D_i$, the 
matter fields $Q_a^{(i)}$ and gauginos living on D$7_i$-branes contribute to the radiative corrections 
through the soft masses~(\ref{eq:softmass}). 
For sake of simplicity, we discuss the D$7$-branes wrapping only the divisor $D_i$ with volume $\tau_i$ 
in the following analysis.

To justify this low-energy effective action, the soft masses and masses of K\"ahler moduli $m_T$ should be 
smaller than those of Kaluza-Klein (KK) modes, stringy modes, complex structure and dilaton moduli, namely 
\begin{align}
m_{T}, m_{3/2} <m_{U}, m_{S} <m_{\rm KK} <m_{\rm st} <M_{\rm Pl},
\end{align}
and their typical masses are given by
\begin{align}
m_{S,U}\simeq N m_{3/2},\qquad
m_{\rm KK}\simeq \frac{\sqrt{\pi}}{{\cal V}^{2/3}}M_{\rm Pl},
\qquad
m_{\rm st}\simeq \frac{\sqrt{\pi}g_s^{1/4}}{{\cal V}^{1/2}}M_{\rm Pl},
\end{align}
with $N>1$ being a parameter determined by three-form fluxes. 
Note that we have taken the string-frame volume ${\cal V}_s={\cal V}g_s^{3/2}\simeq (2\pi R)^6l_s^6$, 
where $R$ is the typical length of CY in string units. 
Since the cutoff scale is a physical quantity in the non-renormalizable theory, 
there exist several options for the cutoff scale. 
For concreteness, we take the cutoff scale as the typical masses of KK modes 
in the following analysis, since there exists $N=2$ supersymmetry above the KK scale. 
The CW potential is then valid below the cutoff scale. 
The string loop corrections to the K\"ahler potential was discussed in Ref.~\cite{Berg:2005yu}, 
and they can also be  interpreted as the 1-loop CW potentials~\cite{Berg:2007wt}. 
However, in this paper, we focus on the situation where sparticle contributions dominate over the 1-loop CW potential.\footnote{The moduli 
stabilization involving the string loop corrections to the K\"ahler potential is discussed 
for CY threefolds with negative Euler number~\cite{Cicoli:2008va} and zero or positive Euler number~\cite{Cicoli:2012fh}.} 
Since the following analysis requires that a relatively large number of sparticle contribute  
to stabilize the CY volume, they could dominate over the string loop corrections to the K\"ahler potential controlled 
by the string coupling.

\subsection{Stabilization of overall volume modulus}
\label{sec:2_2}
In the following analysis, we first consider the large volume region ${\cal V}\rightarrow \infty$, where 
$\tau_i \rightarrow \infty$ with $i=1,2,\cdots, h^{1,1}$ for all of the divisors. (In the LVS, 
some of the divisor volumes are not extremely large in string units~\cite{Balasubramanian:2005zx}.)
When there exist the D-brane instanton effects or gaugino condensation on D$7$-branes, 
the non-perturbative superpotential for the K\"ahler moduli is generated as
\begin{align}
W_{\rm non}=\sum_i A_i e^{-a_i T_i},
\end{align}
where $A_i$ are functions of the complex structure moduli and $a_i=2\pi$ for the brane instanton and $a_i=2\pi/N$ for the 
gaugino condensation on $N$ stacks of D$7$-branes wrapping the divisor with volume $\tau_i$. 
In the moduli space $\tau_i \rightarrow \infty$, these non-perturbative effects are negligible and 
the total scalar potential consists of the leading $\alpha^\prime$-corrections and radiative corrections. 
We come back to these non-perturbative effects in Secs.~\ref{sec:3_3}-~\ref{sec:3_5}. 

Including the radiative corrections due to sfermions and gauginos, we can write the 
scalar potential by
\begin{align}
V_F&=V_{\alpha^\prime}+V_{\rm CW}
\nonumber\\
&=e^{\langle K(S,U)\rangle}\frac{3\xi}{4{\cal V}^3}\frac{|W_0|^2}{M_{\rm Pl}^2}
+
\frac{c_1}{32\pi^2}
\Lambda^2 m_{3/2}^2
+\sum_a\frac{c_b^a}{64\pi^2}m_{a}^4\biggl[ \ln\left(\frac{m_{a}^2}{\Lambda^2}\right)-\frac{1}{2}\biggl]
\nonumber\\
&+\frac{-2c_f-4}{64\pi^2}m_{3/2}^4 \biggl[\ln\left(\frac{m_{3/2}^2}{\Lambda^2}\right)-\frac{1}{2}\biggl]
+\frac{32}{64\pi^2}m_{3/2}^4 \biggl[\ln\left(\frac{4m_{3/2}^2}{\Lambda^2}\right)-\frac{1}{2}\biggl].
\label{eq:VF}
\end{align}
It turns out that the $F$-term scalar potential is a function of only the overall volume ${\cal V}$. 
In the large volume regime ${\cal V}\gg 1$, 
the above scalar potential is further approximated as
\begin{align}
V_F&\simeq e^{\langle K(S,U)\rangle}\frac{3\xi}{4{\cal V}^3}\frac{|W_0|^2}{M_{\rm Pl}^2}
+
\frac{c_1}{32\pi^2}
\Lambda^2 \frac{e^{\langle K(S,U)\rangle}|W_0|^2 }{{\cal V}^2M_{\rm Pl}^4}.
\end{align}

Let us fix the cutoff scale as the KK scale and consider the following units
\begin{align}
&\Lambda= m_{\rm KK} =\frac{\sqrt{\pi}}{{\cal V}^{2/3}}M_{\rm Pl}=1,
\end{align}
and then the dimensional quantities are rewritten as 
\begin{align}
M_{\rm Pl}&=\frac{{\cal V}^{2/3}}{\sqrt{\pi}},
\nonumber\\
W_0&\equiv \hat{W}_0M_{\rm Pl}^3=\hat{W}_0\frac{{\cal V}^2}{\pi^{3/2}}. 
\end{align}
In units of $m_{\rm KK}$, the approximated scalar potential reduces to 
\begin{align}
V_F&\simeq 
e^{\langle K(S,U)\rangle}\frac{|\hat{W}_0|^2}{\pi^2} \biggl[\frac{3\xi}{4{\cal V}^{1/3}}
+
\frac{c_1}{32\pi^2}
\frac{\pi}{{\cal V}^{2/3}}\biggl].
\end{align}

From this scalar potential, 
we aim to find the minimum of ${\cal V}$ in the large volume regime ${\cal V}\gg 1$. 
The extremal condition of modulus field 
$\langle \partial_{{\cal V}}V_F\rangle=0$ in the large volume limit,
\begin{align}
\partial_{{\cal V}}V_F\simeq e^{\langle K(S,U)\rangle}\frac{|\hat{W}_0|^2}{\pi^2}  \biggl[-\frac{\xi}{4{\cal V}^{4/3}}
-
\frac{c_1}{48\pi^2}
\frac{\pi}{{\cal V}^{5/3}}\biggl]=0,
\end{align}
is satisfied at
\begin{align}
&\langle{\cal V}\rangle \simeq 18664 \left(-\frac{c_1/\xi}{10^{3}}\right)^3,
\label{eq:Vnoup}
\end{align}
in string units. 
In particular, when the CY volume is dominated by a single K\"ahler modulus ${\cal V}\simeq \kappa (T+\bar{T})^{3/2}$ with $\kappa$ being a real positive constant, the value of overall volume modulus is evaluated as
\begin{align}
&\langle \tau \rangle\simeq 351.8 \left(-\frac{c_1/\xi}{10^{3}}\right)^{2}\kappa^{-2/3},
\end{align}
where $T\equiv \tau+i\sigma$. 
We find that the ratio $c_1/\xi$ should be negative and larger than $10^{2-3}$ such that 
the K\"ahler moduli space resides in the physical domain.

Furthermore, the positivity of  $\langle \partial_{{\cal V}}\partial_{{\cal V}}V_F\rangle$,
\begin{align}
\partial_{{\cal V}}\partial_{{\cal V}}V_F \simeq 9.14\times 10^{8} e^{\langle K(S,U)\rangle}|\hat{W}_0|^2\frac{\xi^{8}}{c_1^{7}} >0,
\end{align}
requires that $c_1$ should be positive. 
That indicates the negative $\xi$ from Eq.~(\ref{eq:Vnoup}). 
Obviously, the sfermions with $n_i^a=1$ have no contribution in the potential.
Thus, it turns out that the case with $n_i^a=1$ for all sfermions is prohibited, 
because such a case corresponds to the negative $c_1$ 
from Eq.~(\ref{eq:c1}). 
In addition, the potential energy becomes negative at this minimum, namely anti-de Sitter minimum,
\begin{align}
&\langle V_F\rangle \simeq 1.4\times 10^{-3}  e^{\langle K(S,U)\rangle}|\hat{W}_0|^2 \left(-\frac{10^{3}}{c_1/\xi}\right)\xi <0 ,
\end{align}
where $\xi$ is negative.  

At this minimum, the mass scales of typical modes become
\begin{align}
&m_a\simeq M_f\simeq m_{3/2} =\frac{e^{\langle K(S,U)\rangle/2}|\hat{W}_0|}{{\cal V}}M_{\rm Pl}\simeq 
2.1\times 10^{-2} \left(-\frac{10^{3}}{c_1/\xi}\right)e^{\langle K(S,U)\rangle/2}|\hat{W}_0| m_{\rm KK}, 
\nonumber\\
&m_{U,S}=Nm_{3/2}=2.1\times 10^{-2} N \left(-\frac{10^{3}}{c_1/\xi}\right)e^{\langle K(S,U)\rangle/2}|\hat{W}_0| m_{\rm KK}, 
\nonumber\\
&m_{\rm KK}=\frac{\sqrt{\pi}}{{\cal V}^{2/3}}M_{\rm Pl}=1,
\nonumber\\
&m_{\rm st}=\frac{\sqrt{\pi}g_s^{1/4}}{{\cal V}^{1/2}}M_{\rm Pl}
=2.9\left(\frac{g_s}{10^{-1}}\right)^{1/4} \left(-\frac{10^{3}}{c_1/\xi}\right)^{1/2}m_{\rm KK},
\nonumber\\
&M_{\rm Pl}=397\left(-\frac{c_1/\xi}{10^{3}}\right)^{2}m_{\rm KK}.
\label{eq:analytic}
\end{align}
Especially, when the CY volume is dominated by a single K\"ahler modulus, 
the mass scale of canonically-normalized overall K\"ahler modulus is 
\begin{align}
m_{\tau}\simeq 1.1\times 10^{-3}e^{\langle K(S,U)\rangle/2}|\hat{W}_0|\left(-\frac{10^{3}}{c_1/\xi}\right)^{3/2}(-\xi)^{1/2}m_{\rm KK}.
\end{align}


In this way, we find that the volume modulus can be perturbatively stabilized at the anti-de Sitter minimum, where four-dimensional supergravity description is reliable. 
 The non-perturbative effects for the overall K\"ahler 
modulus are suppressed enough by the value of K\"ahler modulus, $e^{-2\pi \tau}\ll 1$, 
but it generates the potential of remaining massless axion as discussed in Sec.~\ref{sec:3_5}. 
However, the negative $\xi$ indicates that we have to consider the CY threefolds with the positive 
Euler number, namely a large number of K\"ahler moduli compared with that of complex structure moduli\footnote{Our 
result is similar to Ref.~\cite{Ciupke:2015msa}, where the higher-derivative corrections are taken into account.}. 
(In the LVS, the negative Euler number is required to realize the large CY volume.) 
Since such CY threefolds with the
positive Euler number account for half of the whole CY threefolds in the sense of mirror symmetry, 
the CYs with the positive Euler number are not so restricted manifolds, but generic. The O7-plane contribution will also help us to change the 
number of CY Euler number in a weak coupling limit~\cite{Minasian:2015bxa}. 
In Secs.~\ref{sec:3_3} and~\ref{sec:3_4}, we discuss the stabilization of other K\"ahler moduli. 
Although for a generic value of $e^{\langle K(S,U)\rangle/2}|\hat{W}_0|$, the overall volume 
modulus can be stabilized at the anti-de Sitter minimum in a way similar to the LVS, 
we require the relatively large $c_1$ to realize the large volume of CY, e.g., 
$c_1\simeq  {\cal O}(10^{2})$ for $|\xi|\simeq {\cal O}(10^{-1})$. 
Such a situation is easily realized by the models with multi-Higgs doublets and singlets via Eq.~(\ref{eq:cb}), 
and such models often appear from concrete string model constructions. (For the string model building, see e.g. Refs.~\cite{Ibanez:2012zz,Blumenhagen:2006ci}.)

Finally, we comment on the case with no leading $\alpha^\prime$-corrections, which corresponds to 
the CY manifold with the vanishing Euler number, i.e., $\chi=0$. 
Then, the scalar potential consists of only the CW potential,
\begin{align}
V_{\rm CW}
&=\frac{c_1}{32\pi^2}
\Lambda^2 m_{3/2}^2
+\frac{c_2}{64\pi^2}m_{3/2}^4\biggl[ \ln\left(\frac{m_{3/2}^2}{\Lambda^2}\right)\biggl]
+\frac{c_3}{64\pi^2}m_{3/2}^4,
\end{align}
with 
\begin{align}
c_2&\equiv \sum_a c_b^a(1-n_i^a)^2-2c_f-4+32,
\nonumber\\
c_3&\equiv \sum_a \biggl[c_b^a(1-n_i^a)^2\left(-\frac{1}{2}+\ln (1-n_i^a)\right)\biggl]+c_f+2+32\left(\ln 4-\frac{1}{2}\right).
\end{align}
It can be rewritten in units of $\Lambda=m_{\rm KK}=1$,
\begin{align}
V_{\rm CW}&\simeq 
e^{\langle K(S,U)\rangle}\frac{|\hat{W}_0|^2}{(32\pi^2)\pi^2} \biggl[
c_1\frac{\pi}{{\cal V}^{2/3}}
+c_2\frac{e^{\langle K(S,U)\rangle}|\hat{W}_0|^2}{{\cal V}^{4/3}}\ln\left(\frac{e^{\langle K(S,U)\rangle/2}\hat{W}_0}{{\cal V}^{1/3}\sqrt{\pi}}\right)
+c_3\frac{e^{\langle K(S,U)\rangle}|\hat{W}_0|^2}{2{\cal V}^{4/3}}
\biggl]
\nonumber\\
&\simeq 
e^{\langle K(S,U)\rangle}\frac{|\hat{W}_0|^2}{(32\pi^2)\pi^2} \biggl[
c_1\frac{\pi}{{\cal V}^{2/3}}
-\frac{c_2}{3}\frac{e^{\langle K(S,U)\rangle}|\hat{W}_0|^2}{{\cal V}^{4/3}}\ln\left({\cal V}\right)
\biggl],
\end{align}
in the large volume regime ${\cal V}\gg 1$ and $e^{\langle K(S,U)\rangle/2}\hat{W}_0\sim {\cal O}(1)$. 
When $c_2$ is almost the same order of $c_1$, we cannot find the minimum 
with a sufficiently large CY volume.

\subsection{Numerical estimation}
\label{sec:2_3}
\begin{figure}[htbp]
  \begin{center}
    \begin{tabular}{c}

      \begin{minipage}{0.5\hsize}
        \begin{center}
          \includegraphics[clip, width=7.0cm]{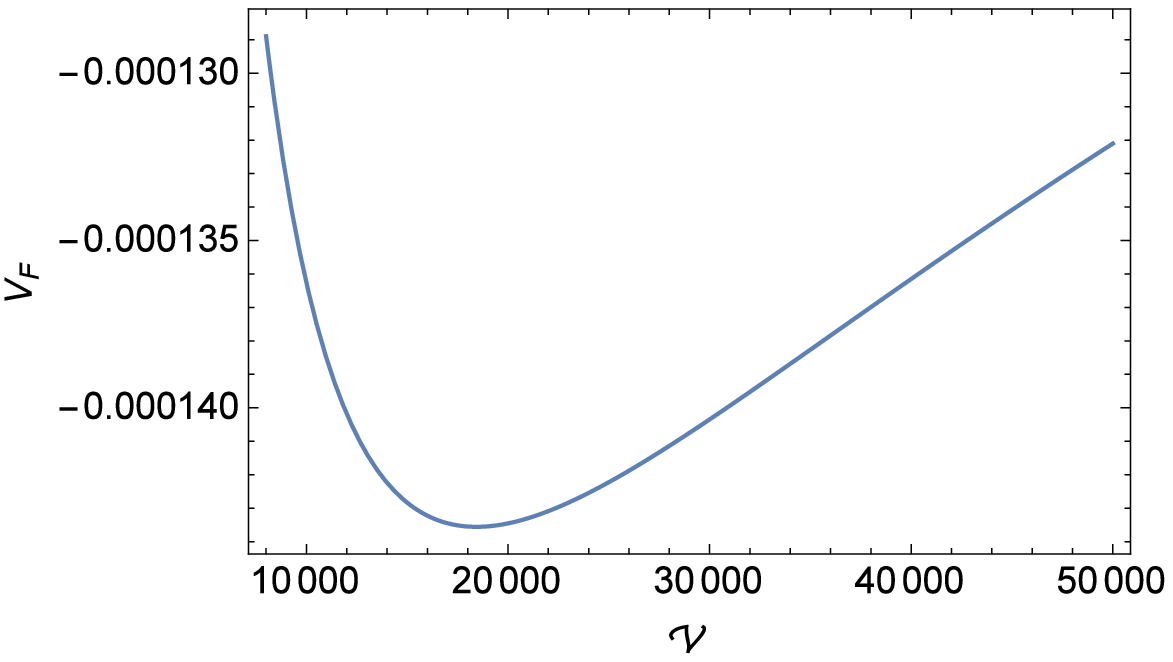}
          \hspace{1.6cm} 
        \end{center}
      \end{minipage}

      \begin{minipage}{0.5\hsize}
        \begin{center}
          \includegraphics[clip, width=7.0cm]{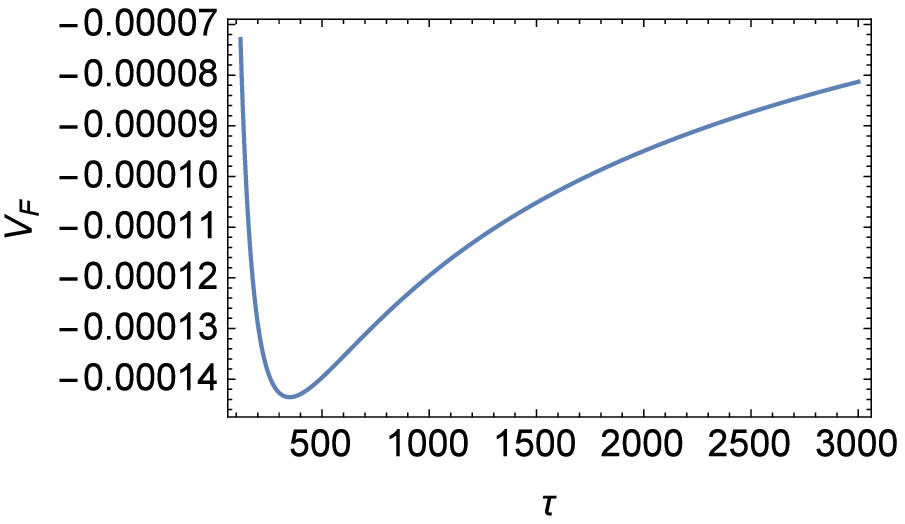}
          \hspace{1.6cm}
        \end{center}
      \end{minipage}

    \end{tabular}
    \caption{The scalar potential~(\ref{eq:VF}) as functions of CY volume ${\cal V}$ and non-canonically normalized modulus $\tau$, where the CY volume is approximated as ${\cal V}=(T+\bar{T})^{3/2}$ with $T=\tau+i\sigma$ 
    in the right panel.}
    \label{fig:poKK}
  \end{center}
\end{figure}

In this section, we numerically analyze the stabilization of moduli fields.
The unapproximated scalar potential of the overall volume modulus~(\ref{eq:VF}) in units of $\Lambda=m_{\rm KK}$ is drawn as functions of ${\cal V}$ 
and $\tau$ in Fig.~\ref{fig:poKK}, where we set the following parameters,
\begin{align}
\xi= -0.1,\qquad
e^{\langle K(S,U)\rangle/2}|\hat{W}_0|= 1,\qquad
n_i^a=0,\qquad
\kappa=1,\qquad
c_b=120,\qquad
c_f=12.
\end{align}

At this vacuum in Fig.~\ref{fig:poKK}, the CY volume is determined by
\begin{align}
\langle{\cal V}\rangle&\simeq 18433,
\label{eq:ReTv}
\end{align}
as shown in the left panel and the vacuum expectation value of overall K\"ahler modulus is
\begin{align}
\langle \tau\rangle&\simeq 348.9,
\label{eq:ReTv}
\end{align}
as shown in the right panel. 
The mass scales of typical modes become
\begin{align}
&m_i=M_f=m_{3/2} =2.1\times 10^{-2}m_{\rm KK},
\nonumber\\
&m_{U,S}=Nm_{3/2}=2.1\times 10^{-2} Nm_{\rm KK},
\nonumber\\
&m_{\rm KK}=1,
\nonumber\\
&m_{\rm st}
=2.9\left(\frac{g_s}{10^{-1}}\right)^{1/4} m_{\rm KK},
\nonumber\\
&M_{\rm Pl}=394m_{\rm KK}.
\label{eq:analytic}
\end{align}
In addition, the mass scale of canonically-normalized overall K\"ahler modulus is 
\begin{align}
m_{\tau}\simeq 3.5\times 10^{-4}m_{\rm KK},
\end{align}
when the volume of CY is dominated by the single K\"ahler modulus $T=\tau+i\sigma$. 
Hence, even when we involve the  next leading terms in the scalar potential, 
which are volume-suppressed, 
the numerical result is in agreement with the analytical one in Sec.~\ref{sec:2_2}.

\section{K\"ahler moduli stabilization with uplifting}
\label{sec:3}
In this section, we introduce the uplifting term to achieve a tiny cosmological constant. 
In particular, we focus on the anti-D3 branes to uplift the anti-de Sitter minimum to the 
de Sitter minimum. 

\subsection{Analytical estimation}
\label{sec:3_1}
When there exist anti-D3 branes at certain moduli space of CY manifold, the uplifting 
term is given by
\begin{align}
V_{\rm up}=\frac{\epsilon}{{\cal V}^2}M_{\rm Pl}^4,
\end{align}
where $\epsilon$ is a real constant suppressed by the warp factor. 
Hence, the total scalar potential is a sum of $V_F$ in Eq.~(\ref{eq:VF}) and $V_{\rm up}$. 

Similar to the analysis in Sec.~\ref{sec:2}, we aim to find the minimum of modulus field analytically. 
In the large volume regime, the above scalar potential in units of $m_{\rm KK}$ is simplified as
\begin{align}
V&=V_F+V_{\rm up}
\simeq e^{\langle K(S,U)\rangle}\frac{|\hat{W}_0|^2}{\pi^2} \biggl[\frac{3\xi}{4{\cal V}^{1/3}}
+
\frac{c_1}{32\pi^2}
\frac{\pi}{{\cal V}^{2/3}}\biggl]
+\frac{\epsilon}{\pi^2}{\cal V}^{2/3},
\end{align}
from which the minimum of overall volume modulus is determined by solving $\partial_{\cal V}V=0$, 
\begin{align}
\langle{\cal V}\rangle&\simeq 5530 \left(-\frac{c_1/\xi}{10^{3}}\right)^3,
\label{eq:CYup}
\end{align}
and $\epsilon$ is chosen such that $V\simeq 0$,
\begin{align}
\epsilon&\simeq 3.4\times 10^{-5}e^{\langle K(S,U)\rangle/2}|\hat{W}_0|^2 (-\xi)\left(-\frac{10^{3}}{c_1/\xi}\right)^3.
\end{align}
In particular, when the CY volume is dominated by a single K\"ahler modulus ${\cal V}\simeq \kappa (T+\bar{T})^{3/2}$ with $\kappa$ being a real positive constant, the modulus value at this minimum is evaluated as
\begin{align}
\langle\tau \rangle&\simeq 156 \left(-\frac{c_1/\xi}{10^{3}}\right)^{2}\kappa^{-2/3}.
\end{align}
It turns out that the deviation from the SUSY-breaking anti-de Sitter minimum to the uplifted minimum 
is estimated as
\begin{align}
\frac{\langle\tau \rangle|_{\rm anti-de~Sitter}-\langle\tau \rangle|_{\rm de~Sitter}}{\langle\tau \rangle|_{\rm anti-de~Sitter}}&\simeq 0.56.
\end{align}
Our scalar potential still resides in the supergravity-controlled regime, 
since the mass scales of typical modes have a desirable hierarchical structure,
\begin{align}
&m_a\simeq M_f\simeq m_{3/2} =\frac{e^{\langle K(S,U)\rangle/2}|\hat{W}_0|}{\langle {\cal V} \rangle}M_{\rm Pl}\simeq 
3.2\times 10^{-2} \left(-\frac{10^{3}}{c_1/\xi}\right)e^{\langle K(S,U)\rangle/2}|\hat{W}_0| m_{\rm KK}, 
\nonumber\\
&m_{U,S}=Nm_{3/2}=3.2\times 10^{-2} N \left(-\frac{10^{3}}{c_1/\xi}\right)e^{\langle K(S,U)\rangle/2}|\hat{W}_0| m_{\rm KK}, 
\nonumber\\
&m_{\rm KK}=\frac{\sqrt{\pi}}{\langle{\cal V} \rangle^{2/3}}M_{\rm Pl}=1,
\nonumber\\
&m_{\rm st}=\frac{\sqrt{\pi}g_s^{1/4}}{\langle{\cal V} \rangle^{1/2}}M_{\rm Pl}
=2.4\left(\frac{g_s}{10^{-1}}\right)^{1/4} \left(-\frac{10^{3}}{c_1/\xi}\right)^{1/2}m_{\rm KK},
\nonumber\\
&M_{\rm Pl}=176\left(-\frac{c_1/\xi}{10^{3}}\right)^{2}m_{\rm KK}.
\label{eq:analyticup}
\end{align}
Furthermore, when the CY volume is dominated by a single K\"ahler modulus, 
the mass scale of canonically-normalized overall K\"ahler modulus is 
\begin{align}
m_{\tau}\simeq 3.5\times 10^{-3}e^{\langle K(S,U)\rangle/2}|\hat{W}_0|\left(-\frac{10^{3}}{c_1/\xi}\right)^{3/2}(-\xi)^{1/2}m_{\rm KK}.
\end{align}

We find that the volume modulus is still stabilized at the supergravity-reliable de Sitter minimum, 
even after the anti-de Sitter minimum is uplifted to the de Sitter vacuum. 
By setting $e^{\langle K(S,U)\rangle/2}|\hat{W}_0|=1$, $g_s=0.1$, and $\xi=-0.1$, we list the volume of CY,  the typical scales of gravitino, moduli fields, KK and stringy modes for several values of $c_1$ in Table~\ref{tab:1}.

\begin{table}[htb]
  \begin{center}
    \begin{tabular}{|c|c|c|c|} \hline
      Scale & $c_1=50$ & $c_1=100$ & $c_1=1000$ \\ \hline \hline
      ${\cal V}$ & $691$ & $5530$ & $5.5\times 10^6$\\ \hline 
      $\tau$ & $39\kappa^{-2/3}$ & $156\kappa^{-2/3}$ & $1.6\times 10^4\kappa^{-2/3}$\\ \hline 
      $m_\tau$[GeV] & $1.7\times 10^{14}$ & $1.5\times 10^{13}$ & $4.8\times 10^{9}$\\ \hline 
      $m_{3/2}$[GeV] & $3.5\times 10^{15}$ & $4.4\times 10^{14}$ & $4.4\times 10^{11}$\\ \hline 
      $m_{U,S}$[GeV] & $3.5N\times 10^{15}$ & $4.4N\times 10^{14}$ & $4.4N\times 10^{11}$\\ \hline 
      $m_{\rm KK}$[GeV] & $5.5\times 10^{16}$ & $1.4\times 10^{16}$  & $1.4\times 10^{14}$ \\ \hline 
      $m_{\rm st}$[GeV] & $9.1\times 10^{16}$ & $3.2\times 10^{16}$ & $1.0\times 10^{15}$ \\ \hline 
      $M_{\rm Pl}$[GeV] & $2.4\times 10^{18}$ & $2.4\times 10^{18}$ & $2.4\times 10^{18}$ \\ \hline 
    \end{tabular}
  \end{center}
    \caption{The volume of CY (${\cal V}$) and divisor ($\tau$), and typical scales of several modes, setting $e^{\langle K(S,U)\rangle/2}|\hat{W}_0|=1$, $g_s=0.1$, and $\xi=-0.1$.}
    \label{tab:1}
\end{table}

\subsection{Numerical estimation}
\label{sec:3_2}
\begin{figure}[htbp]
  \begin{center}
    \begin{tabular}{c}

      \begin{minipage}{0.5\hsize}
        \begin{center}
          \includegraphics[clip, width=7.0cm]{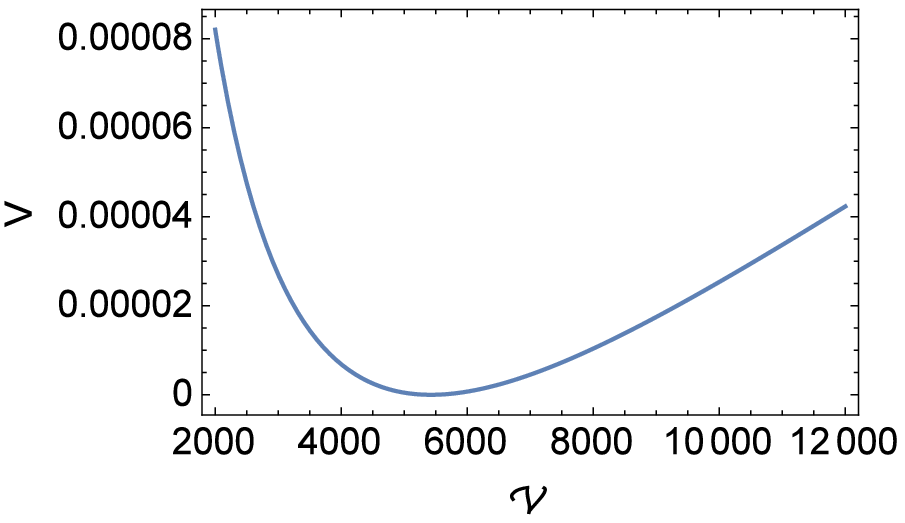}
          \hspace{1.6cm} 
        \end{center}
      \end{minipage}

      \begin{minipage}{0.5\hsize}
        \begin{center}
          \includegraphics[clip, width=7.0cm]{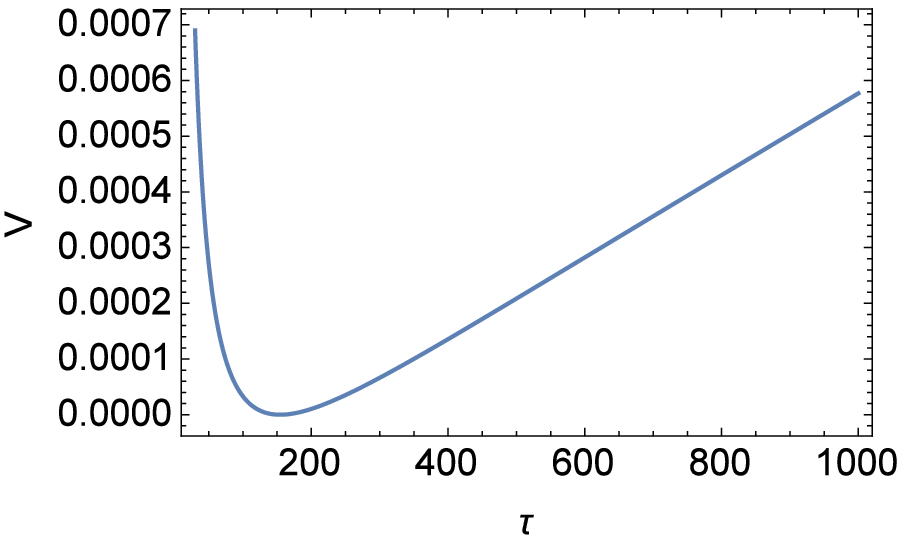}
          \hspace{1.6cm}
        \end{center}
      \end{minipage}

    \end{tabular}
    \caption{The scalar potential in units of $\Lambda=m_{\rm KK}$ as functions of CY volume ${\cal V}$ and non-canonically normalized modulus $\tau$, where the CY volume is approximated as ${\cal V}=(T+\bar{T})^{3/2}$ with $T=\tau+i\sigma$ 
    in the right panel.}
    \label{fig:poKKup}
  \end{center}
\end{figure}

Next, let us numerically estimate the unapproximated scalar potential $V=V_F+V_{\rm up}$ 
setting the same parameters as ones in Sec.~\ref{sec:2_3} and
\begin{align}
\epsilon\simeq 3.437\times 10^{-6},
\end{align}
under which the scalar potential with the uplifting term is drawn in Fig.~\ref{fig:poKKup}.

In the left panel in Fig.~\ref{fig:poKKup}, the CY volume is determined as
\begin{align}
\langle{\cal V}\rangle&\simeq 5424,
\label{eq:ReTv}
\end{align}
and in the right panel, the vacuum expectation value of overall K\"ahler modulus is
\begin{align}
\langle \tau \rangle&\simeq 154.4.
\label{eq:ReTv}
\end{align}
The mass scales of typical modes become
\begin{align}
&m_i=M=m_{3/2} =3.2\times 10^{-2}m_{\rm KK},
\nonumber\\
&m_{U,S}=Nm_{3/2}=3.2\times 10^{-2} Nm_{\rm KK},
\nonumber\\
&m_{\rm KK}=1,
\nonumber\\
&m_{\rm st}
=2.4\left(\frac{g_s}{10^{-1}}\right)^{1/4} m_{\rm KK},
\nonumber\\
&M_{\rm Pl}=174m_{\rm KK},
\label{eq:analytic}
\end{align}
and in addition the mass scale of canonically-normalized overall K\"ahler modulus is 
\begin{align}
m_{\tau}\simeq 1.1\times 10^{-3}m_{\rm KK}.
\end{align}
Hence, even when we involve the next leading terms in the scalar potential, 
which are volume-suppressed,  
the numerical result is in agreement with the analytical one.

\subsection{Inclusion of the $D$-terms}
\label{sec:3_3}
So far, we have not considered the non-perturbative effects for the K\"ahler moduli, 
since those are negligible in the large volume regime ${\cal V}\rightarrow \infty$, where $\tau_i\rightarrow \infty$ 
for all  the volumes of divisors. 
To stabilize the K\"ahler moduli fields expect for the overall volume modulus, we 
consider the $D$-term potential induced by the Fayet-Iliopoulos (FI) term~\cite{Jockers:2005zy}. 
When the anomalous $U(1)$ symmetries appear on D$7_i$-branes wrapping 
the divisors $\tau_i$, the $D$-term potential is written by
\begin{align}
V_{D}^{{\rm D7}}=\sum_i \frac{1}{{\rm Re}(f_{{\rm D7}_i})} \left(q_{T_i}^{({\rm D7}_i)}K_{T_i}-\sum_m q_{\varphi^m}^{({\rm D7}_i)}|\phi_m|^2\right)^2,
\end{align}
where $f_{{\rm D7}_i}$ are the gauge kinetic functions of D$7_i$-branes, $q_{T_i}^{({\rm D7}_i)}$ represent the 
gauge fluxes and $\phi_m$ with U(1) charges $q_{\varphi^m}^{({\rm D7}_i)}$ are the canonically normalized matter fields living on D$7_i$-branes. 
In addition to these D$7$-brane contributions, anomalous U(1) symmetries on the fractional D$3$-branes 
located at the singularities of CY also induce the $D$-terms
\begin{align}
V_{D}^{{\rm D3}}=\sum_k \frac{1}{{\rm Re}(f_{{\rm D3}_k})} \left(q_{T_k}^{({\rm D3}_k)}K_{T_k}-\sum_m q_{\varphi^m}^{({\rm D3}_k)}|\varphi_m|^2\right)^2,
\end{align}
where $f_{{\rm D3}_k}$ are the gauge kinetic functions of D$3_k$-branes with $k$ being the number of singularities, 
$q_{T_k}^{({\rm D3}_k)}$ represent the U(1) charges of moduli fields and $\varphi_m$ with U(1) charges $q_{\varphi^m}^{({\rm D3})}$ 
are the canonically normalized matter fields living on the fractional D$3_k$-branes. 
In both cases, the anomalous U(1) gauge bosons eat the linear combination of string axions through 
the St\"uckelberg couplings and become massive at the compactification scale. 
In order not to spoil the stabilization of $F$-term potential, we require the vanishing $D$-term potential, $V_{D}^{{\rm D7}}=V_{D}^{{\rm D3}}=0$, namely 
\begin{align}
q_{T_i}^{({\rm D7}_i)}K_{T_i}=\sum_m q_{\varphi^m}^{({\rm D7}_i)}|\phi_m|^2, \qquad
q_{T_k}^{({\rm D3}_k)}K_{T_k}=\sum_m q_{\varphi^m}^{({\rm D3}_k)}|\varphi_m|^2,
\end{align}
which fix the linear combination of closed string moduli and matter fields (open string moduli). 
In addition, we consider the $F$-term potential of matter fields, 
\begin{align}
V_F^{{\rm matter}}=m_{\Phi}^2|\Phi|^2+A_\Phi|\Phi|^3+\lambda_\Phi|\Phi|^4,
\end{align}
where $\Phi$ denotes $\phi_m$ and $\varphi_m$, $m_{\Phi}^2,A_\Phi,\lambda_\Phi$ are $T_i$-dependent functions. 
When $\Phi$ develops a nonvanishing value, the $D$-term and $F$-term contributions are possible to 
stabilize all the moduli fields except for the overall volume modulus, although it depends on 
a topology of CY manifold. 
We leave an explicit moduli stabilization for a future work, since it is difficult to analyze the topology of CY 
threefold with huge $h^{1,1}$. (For the model building in the LVS with small $h^{1,1}$, see Ref.~\cite{Cicoli:2013cha}.)

\subsection{Inclusion of the non-perturbative effects}
\label{sec:3_4}
In contrast to the previous section, we study alternative stabilization scenario for 
K\"ahler moduli other than the overall volume modulus.
We consider the large volume regime ${\cal V}\rightarrow \infty$, 
but the volume of one divisor $\tau_s$ is not extremely larger than the other K\"ahler moduli 
$\tau_i$, namely $\tau_i \gg \tau_s > 1$ in string units. 
Note that we focus on the K\"ahler cone where $\tau_s$ is larger than the string length to 
suppress the KK and stringy corrections. 

Let us assume the non-pertrubative superpotential with respect to $\tau_s$, 
\begin{align}
W=W_0+A_s e^{-a_sT_s},
\label{eq:W}
\end{align}
where $W_0=\langle W_{\rm flux}\rangle$, $A_s$ and $a_s$ are the constants 
depending on the origin of non-perturbative effects, e.g., 
$a_s=2\pi$ for the brane instanton and $a_s=2\pi/N$ for the 
gaugino condensation on $N$ stacks of D$7$-branes wrapping the divisor with volume $\tau_s$. 
To simplify the following analysis, we take both the $W_0$ and $A_s$ as real constants. 
From the K\"ahler potential~(\ref{eq:K}) and superpotential~(\ref{eq:W}), the 
$F$-term scalar potential is simply given by
\begin{align}
V_FM_{\rm Pl}^2&=e^{K}\left(K^{T_i\bar{T}_j}D_{T_i}WD_{\bar{T}_j}\bar{W}-3\right)|W|^2
\nonumber\\
&\simeq \frac{e^{K(S,U)}}{{\cal V}^2}\biggl[K^{T_s\bar{T}_s}a_s^2|A_s|^2e^{-2a_s\tau_s}
-K^{T_s\bar{T}_j}(\partial_{\bar{T}_j}K)a_s A_s e^{-a_sT_s}\bar{W}
-K^{T_i\bar{T}_s}(\partial_{T_i}K)a_s \bar{A}_s e^{-a_s\bar{T}_s}W
\biggl]
\nonumber\\
&+\frac{3\xi}{4{\cal V}^3}e^{K(S,U)}|W|^2 +{\cal O}\left(\frac{1}{{\cal V}^4}\right)
\nonumber\\
&\simeq e^{K(S,U)}\biggl[\frac{4}{{\cal V}}a_s^2|A_s|^2(-\kappa_{ssi}t^i)e^{-2a_s\tau_s}
-\frac{4}{{\cal V}^2}a_s\tau_s|A_s W|e^{-a_s\tau_s}
+\frac{3\xi}{4{\cal V}^3}|W|^2\biggl] +{\cal O}\left(\frac{1}{{\cal V}^4}\right),
\label{eq:FnonLVS}
\end{align}
in the large volume regime, where $D_{T_i}W=W_{T^i}+K_{T_i}W$ with $W_{T^i}=\partial_{T^i}W$ and 
the CY volume is taken as ${\cal V}=\sum_{ijk}\frac{\kappa_{ijk}}{6}t^it^jt^k$ 
with $\kappa_{ijk}$ being the intersection numbers among the two-cycles $t^i$ of CY. 
Now, the imaginary part of $T_s$ is set as its minimum and we use $K^{T_i\bar{T}_j}K_{\bar{T}_j}=-2\tau_i +{\cal O}({\cal V}^{-1})$ 
and $K^{T_i\bar{T}_j}= -4{\cal V}\kappa_{ijk}t^k +4\tau_i\tau_j +{\cal O}({\cal V}^{-1})$~\cite{Bobkov:2004cy}, 
where the divisor volume is defined as $\tau_i=\frac{1}{2}\kappa_{ijk}t^jt^k$. 
Note that the first term in Eq.~(\ref{eq:FnonLVS}) is positive, since it is originating from 
the positive definite term. 
The simplified above scalar potential is a well-known form as discussed in the LVS, where 
the coefficient of $\alpha^\prime$-correction $\xi$ is positive, namely $h^{2,1}>h^{1,1}>1$. 
In LVS, the ``small'' modulus $\tau_s$ and volume modulus ${\cal V}$ are stabilized at 
\begin{align}
{\cal V}\sim e^{a_s\tau_s},\qquad \tau_s\sim \xi^{2/3}, 
\end{align} 
because of the positivity of $\xi$.

On the contrary, in this paper, we proceed to discuss the opposite sign of $\xi$, namely $h^{1,1}\geq h^{2,1}>1$. 
The scalar potential is given by a sum of the $F$-term and leading radiative corrections,
\begin{align}
V&\simeq e^{K(S,U)}M_{\rm Pl}^4\biggl[\frac{4}{{\cal V}}a_s^2|\hat{A}_s|^2(-\kappa_{ssi}t^i)e^{-2a_s\tau_s}
-\frac{4}{{\cal V}^2}a_s\tau_s|\hat{A}_s \hat{W}|e^{-a_s\tau_s}
+\frac{3\xi}{4{\cal V}^3}|\hat{W}|^2 
+\frac{c_1}{32\pi^2}\Lambda^2
\frac{|\hat{W}|^2 }{{\cal V}^{2}M_{\rm Pl}^2}\biggl]
\nonumber\\
&+{\cal O}\left(\frac{1}{{\cal V}^4}\right),
\label{eq:Fnon}
\end{align}
where $A_s=\hat{A}_sM_{\rm Pl}^3$ and $W=\hat{W}M_{\rm Pl}^3$. 
In units of $\Lambda=m_{\rm KK}=\sqrt{\pi}M_{\rm Pl}/{\cal V}^{2/3}=1$, 
the scalar potential reduces to 
\begin{align}
V&\simeq \frac{e^{K(S,U)}}{\pi^2}\biggl[4{\cal V}^{5/3}a_s^2|\hat{A}_s|^2(-\kappa_{ssi}t^i)e^{-2a_s\tau_s}
-4{\cal V}^{2/3}a_s\tau_s|\hat{A}_s \hat{W}|e^{-a_s\tau_s}
+\frac{3\xi}{4{\cal V}^{1/3}}|\hat{W}|^2 
+\frac{c_1}{32\pi}
\frac{|\hat{W}|^2 }{{\cal V}^{2/3}}\biggl]\nonumber\\
&
+{\cal O}\left(\frac{1}{{\cal V}^{4/3}}\right).
\end{align}
 
The extremal condition of $\tau_s$ reads
\begin{align}
\frac{\partial V}{\partial \tau_s}&\simeq 
4{\cal V}^{5/3}a_s^2|\hat{A}_s|^2e^{-2a_s\tau_s}\left[
\frac{\partial t^j}{\partial \tau_s}(-\kappa_{ssj})-2a_s(-\kappa_{ssi}t^i)\right]
-4{\cal V}^{2/3}a_s|\hat{A}_s \hat{W}_0|e^{-a_s\tau_s} \left[1-a_s\tau_s\right]
\nonumber\\
&\simeq 
4{\cal V}^{5/3}a_se^{-2a_s\tau_s}\biggl[-2a_s^2|\hat{A}_s|^2(-\kappa_{ssi}t^i)
+\frac{e^{a_s\tau_s}}{{\cal V}}|\hat{A}_s\hat{W}_0| a_s\tau_s\biggl]=0,
\end{align}
with $W_0=\hat{W}_0M_{\rm Pl}^3$, where we use $\partial t^j/\partial \tau_s<1$ 
and $a_s\tau_s \gg 1$ to suppress the higher instanton effects. 
In this way, we obtain
\begin{align}
\frac{e^{a_s\tau_s}}{{\cal V}}|\hat{A}_s\hat{W}_0| a_s\tau_s=2a_s^2|\hat{A}_s|^2(-\kappa_{ssi}t^i).
\label{eq:taus3_4}
\end{align}
Another extremal condition of ${\cal V}$ gives rise to
\begin{align}
\frac{\partial V}{\partial {\cal V}}&\simeq 
\frac{20{\cal V}^{2/3}}{3}a_s^2|\hat{A}_s|^2(-\kappa_{ssi}t^i)e^{-2a_s\tau_s}
-\frac{8a_s\tau_s}{3{\cal V}^{1/3}}|\hat{A}_s \hat{W}_0|e^{-a_s\tau_s}
-\frac{\xi}{4{\cal V}^{4/3}}|\hat{W}_0|^2 
-\frac{c_1}{48\pi}
\frac{|\hat{W}_0|^2 }{{\cal V}^{5/3}}
\nonumber\\
&\simeq 
\frac{2a_s\tau_s}{3{\cal V}^{1/3}}|\hat{A}_s \hat{W}_0|e^{-a_s\tau_s}
-\frac{\xi}{4{\cal V}^{4/3}}|\hat{W}_0|^2 
-\frac{c_1}{48\pi}
\frac{|\hat{W}_0|^2 }{{\cal V}^{5/3}}
\nonumber\\
&\simeq 
\frac{1}{3{\cal V}^{3/4}}\frac{(\tau_s)^2}{(-\kappa_{ssi}t^i)}|\hat{W}_0|^2
-\frac{\xi}{4{\cal V}^{4/3}}|\hat{W}_0|^2 
-\frac{c_1}{48\pi}
\frac{|\hat{W}_0|^2 }{{\cal V}^{5/3}}
\nonumber\\
&\equiv 
-\frac{\hat{\xi}}{4{\cal V}^{4/3}}|\hat{W}_0|^2 
-\frac{c_1}{48\pi}
\frac{|\hat{W}_0|^2 }{{\cal V}^{5/3}},
\end{align}
where 
\begin{align}
\hat{\xi}\equiv \xi-\frac{4}{3}\frac{(\tau_s)^2}{(-\kappa_{ssi}t^i)}.
\label{eq:hatxi}
\end{align}
Consequently, the volume modulus is stabilized at
\begin{align}
&{\cal V}\simeq 18664 \left(-\frac{c_1/\hat{\xi}}{10^{3}}\right)^3,
\label{eq:V3_4}
\end{align}
by replacing $\xi$ of Eq.~(\ref{eq:Vnoup}) into $\hat{\xi}$. 
Although the $\hat{\xi}$ depends on the topology of CY manifold as in Eq.~(\ref{eq:Vnoup}), 
the stabilization of volume modulus ${\cal V}$ and $\tau_s$ is achieved inside the K\"ahler cone 
with ${\cal V} \gg \tau_s >1$. 
Indeed, when $\kappa_{ssi}t^i\sim -\tau_s^{1/2}$, $\hat{A}_s=\hat{W}_0=1$, $c_1=200$, $\xi =-10^{-1}$ and $a_s=2\pi$, 
Eqs.~(\ref{eq:taus3_4}) and.~(\ref{eq:V3_4}) give
\begin{align}
{\cal V}\simeq 4098,\qquad
\tau_s \simeq 1.69.
\label{eq:Vtaus}
\end{align}
Furthermore, even if the Euler number of CY is vanishing, i.e., $\xi=0$, both the moduli fields can be 
stabilized at ${\cal V}\simeq 3444, \tau_s \simeq 1.66$, where we set the same parameters of Eq.~(\ref{eq:Vtaus}) 
except for $\xi$. 
Following the same procedure of Sec.~\ref{sec:3_1}, we can achieve the de Sitter vacuum by including  
 the anti-D$3$ branes.

\subsection{Ultralight axion}
\label{sec:3_5}
Finally, we take a closer look at the mass of axion associated with the overall volume modulus. 
To simplify our analysis, we take into account the CY volume dominated 
by the single K\"ahler modulus $T$, i.e., ${\cal V}=\kappa (T+\bar{T})^{3/2}$ with $\kappa$ 
being a positive real constant. 
When there exist the D-brane instanton effects or gaugino condensation on D$7$-branes, 
the axion potential can be extracted from the superpotential
\begin{align}
W=W_0+A^{(T)} e^{-\frac{2\pi}{n}T},
\end{align}
where $W_0=\langle W_{\rm flux}\rangle$ and $A^{(T)}$ are assumed to be real constants, 
and $n=1$ for the brane instanton and $n=N$ for the 
gaugino condensation on $N$ stacks of D$7$-branes wrapping the divisor with volume $\tau$. 
Such a non-perturbative superpotential term generates the axion potential,
\begin{align}
V&=e^{K/M_{\rm Pl}^2}K^{T\bar{T}}K_T\left(\frac{W_T \bar{W}}{M_{\rm Pl}^2} +\frac{\bar{W}_{\bar{T}} W}{M_{\rm Pl}^2}\right)
\nonumber\\
&\simeq \frac{e^{K(S,U)}A^{(T)} W_0}{{\cal V}^2M_{\rm Pl}^2}\frac{(T+\bar{T})^2}{3}\frac{3}{T+\bar{T}}
\frac{2\pi}{n} \left(e^{-\frac{2\pi}{n} T} +e^{-\frac{2\pi}{n} \bar{T}}\right)
\nonumber\\
&\simeq \frac{e^{K(S,U)}A^{(T)} W_0}{{\cal V}^2M_{\rm Pl}^2} \left(\frac{{\cal V}}{\kappa}\right)^{2/3}
\frac{4\pi}{n} e^{-\frac{2\pi}{n} \tau}\cos \left(\frac{2\pi}{n}\sigma \right),
\end{align}
where $T=\tau +i\sigma$. 
After canonically normalizing the axion $\theta=\sqrt{2K_{T\bar{T}}}\sigma$, we obtain the tiny mass of axion 
and its decay constant,
\begin{align}
m_\theta&\simeq m_{3/2}\frac{(|A^{(T)}|/M_{\rm Pl}^3)^{1/2}}{\sqrt{3}\kappa}
\left(\frac{2\pi}{n}\right)^{3/2}\left(\frac{{\cal V}}{\kappa}\right) e^{-\frac{\pi}{2n} \left(\frac{{\cal V}}{\kappa}\right)^{2/3}},
\nonumber\\
f_\theta&=\frac{n\sqrt{2K_{T\bar{T}}}}{2\pi}=\frac{\sqrt{6}n}{2\pi}\left(\frac{\kappa}{{\cal V}}\right)^{2/3}M_{\rm Pl},
\end{align}
where the CY volume and gravitino mass are given in Eqs.~(\ref{eq:CYup}) and (\ref{eq:analyticup}). 
Thus, large volume of CY results in the ultralight axion associated with the volume modulus in a way similar to the LVS. 

Although the axion mass highly depends on the value of $n$, 
we list the typical axion mass in Table~\ref{tab:2} setting the same parameters 
of Sec.~\ref{sec:2_3} and
\begin{align}
\kappa=1,\qquad
|A^{(T)}|=M_{\rm Pl}^3,\qquad
|c_1/\xi|=10^3.
\end{align}
It is interesting to discuss the astrophysical and cosmological physics of such an ultralight axion, 
which will be studied in a separate work. 


\begin{table}[htb]
  \begin{center}
    \begin{tabular}{|c|c|c|c|c|c|} \hline
      Scale & $n=1$ & $n=3$ & $n=5$  & $n=7$ & $n=9$ \\ \hline 
      $m_\theta/m_{3/2}$ & $2.3\times 10^{-209}$ & $7.4\times 10^{-68}$ & $9.7\times 10^{-40}$ & $9.1\times 10^{-28}$ & $3.7\times 10^{-21}$\\ \hline 
      $f_\theta$[GeV]  & $3\times 10^{15}$ & $9\times 10^{15}$ & $1.5\times 10^{16}$  & $2\times 10^{16}$ & $2.7\times 10^{16}$\\ \hline 
    \end{tabular}
  \end{center}
    \caption{Axion mass in units of gravitino mass and its decay constant.}
    \label{tab:2}
\end{table}

\section{Conclusion}
\label{sec:con}
We have discussed the stabilization of K\"ahler moduli using the leading $\alpha^\prime$-corrections and 
the radiative corrections due to the sparticles within the framework of the type IIB superstring theory on 
Calabi-Yau orientifolds with D$7$-branes. 
When all the volumes of the divisors in CY threefold are sufficiently 
large in string units, the non-perturbative effects for the K\"ahler moduli are 
suppressed enough compared with $\alpha^\prime$-corrections and the radiative corrections. 
We find that these perturbative corrections give rise to the stabilization of the overall K\"ahler modulus 
for a general class of CY threefolds, only if the Euler number of CY is positive. 
Since the volume of CY scales as the number of sparticles, we require that the relatively large number of 
sparticles contributes to the Coleman-Weinberg potential through the radiative corrections to achieve the large 
CY volume. Such a large number of sparticle contributions could dominate over the string loop corrections to 
the K\"ahler potential discussed in Ref.~\cite{Berg:2005yu} where one needs a certain amount of fine-tuning 
of the complex structure moduli to realize the large volume of CY. 
Furthermore, our scenario of moduli stabilization does not require the tuning of flux-induced superpotential in contrast to the KKLT scenario. 
The Kaluza-Klein and stringy modes are sufficiently heavier than  K\"ahler moduli and 
sparticles. In this reason, the low-energy effective action is controllable in the four-dimensional $N=1$ supergravity. 
The vacuum in our scenario of moduli stabilization is the ant-de Sitter minimum. 
The structure of moduli stabilization is still maintained even after we uplift the 
anti-de Sitter minimum to de Sitter vacuum by introducing anti-D3 branes. 
Note that a requirement of the positive Euler number is different from the conventional LARGE volume scenario, 
but CY threefolds with the positive Euler number are accounted for half of the CY threefolds in the sense of mirror symmetry.

However, the positive Euler number of CY threefold indicates that we have to take into account 
a lot of K\"ahler moduli compared with the complex structure moduli. 
We expect that the $D$-term contribution from fractional D$3$-branes and magnetized D$7$-branes 
would lead to the stabilization of these K\"ahler moduli except for the overall volume modulus as discussed 
in Sec.~\ref{sec:3_3}. The non-perturbative effects also allow us to stabilize some K\"ahler moduli 
without spoiling the stabilization of volume modulus as shown in Sec.~\ref{sec:3_4}. 
In this moduli stabilization scenario, the axion associated with the volume modulus remains light and 
it could be a target of astrophysical and cosmological observations. 
It is interesting to show explicit moduli stabilization in a detailed setup, but we leave it for a future work.

\section*{Acknowledgments}
T.~K. was is supported in part by MEXT KAKENHI Grant Number JP17H05395 
and JSPS KAKENHI Grant Number JP26247042. 
H.~O. was supported in part by JSPS KAKENHI Grant Number JP17K14303.


\end{document}